\documentclass{elsart}

\usepackage{graphicx}

\usepackage{amssymb}

\begin{document}

\begin{frontmatter}



\title{XMM-Newton temperature maps for five intermediate redshift
clusters of galaxies }


\author{Florence Durret}
\address{IAP, 98bis Bd Arago, 75014 Paris, France}
\author{Gast\~ao B. Lima Neto}
\address{IAG/USP, R. do Mat\~ao 1226, 
05508-090 S\~ao Paulo/SP, Brazil}

\begin{abstract}

We have analyzed XMM-Newton archive data for five clusters of galaxies
(redshifts 0.223 to 0.313) covering a wide range of dynamical states,
from relaxed objects to clusters undergoing several mergers. We
present here temperature maps of the X-ray gas together with a
preliminary interpretation of the formation history of these clusters.

\end{abstract}

\end{frontmatter}




\section{Introduction}

There is clear evidence that clusters of galaxies are young 
structures, as predicted by the hierarchical scenario of $\Lambda$CDM cosmology
(e.g.  Kauffmann \& White, 1993; West et al., 1995). The study of cluster
formation is paramount to our understanding of structure formation in the
universe and clusters can effectively be used as cosmological probes
(e.g., to measure the dark energy equation of state).

XMM-Newton now allows us to map with unprecedented sensitivity several
physical parameters of the X-ray gas in clusters of galaxies.
Temperature maps have revealed that even when the X-ray emission
appears smooth, evidence for past or present merging events can be
found; they are therefore precious tools to trace the individual
histories of cluster formation and evolution.

We have analyzed XMM-Newton MOS1, MOS2 and pn data for a sample of
five medium redshift ($z=0.223$--0.313, corresponding to a mean lookback time
of $\sim 3$~Gyr) clusters covering a range of merging histories, derived
temperature maps for the X-ray gas and discuss 
merging scenarios for these clusters.

\section{The data and temperature maps}

\begin{table*}
\caption{Summary of the data}
\begin{center}
\
\begin{tabular}{lrrrrrr}
\hline
Cluster    & Obs. nb  & Initial exp. time (s) & Clean exp. time (s) & redshift \\[-5pt]
           &          & MOS1/MOS2/PN          & MOS1/MOS2/PN        &         \\
\hline
Cl 2137-2353 & 0008830101 & 21833/21834/17567 & ~9714/~9716/~6039 & 0.313  \\[-5pt]
Abell 68     & 0084230201 & 29243/29259/22295 & 23750/22365/15573 & 0.255  \\[-5pt]
Abell 2390   & 0111270101 & 22223/22227/17985 & ~9381/~8861/~6337 & 0.228  \\[-5pt]
Abell 1763   & 0084230901 & 25881/25898/19451 & 11996/12335/~8444 & 0.223  \\[-5pt]
Abell 2744   & 0042340101 & 17390/17420/11581 & 13673/13627/~9174 & 0.306  \\
\hline
\end{tabular}
\end{center}
\label{table:data}
\end{table*}

\begin{figure}
\includegraphics[width=0.9\textwidth]{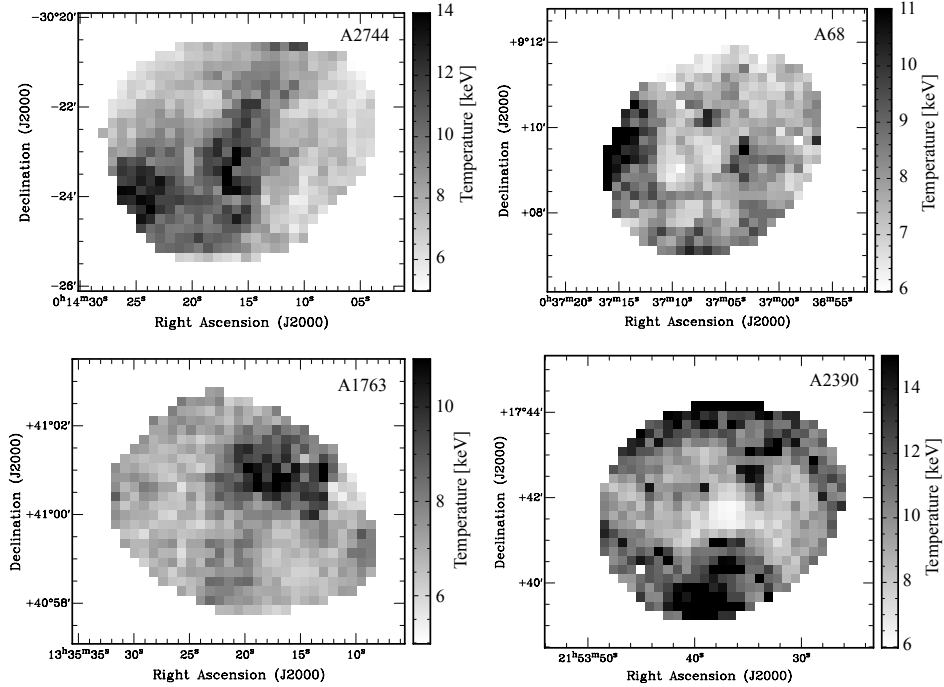}
\caption[]{Temperature maps of Abell~68, 1763, 2390, and 2744.}
\label{fig:fig1}
\end{figure}

\begin{figure}
    \centering
\includegraphics[width=0.4\textwidth]{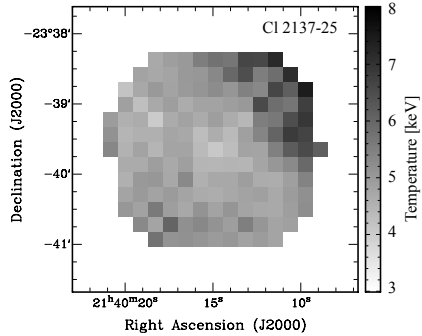}
\caption[]{Temperature map of Cl 2137-2353.}
\label{fig:fig2}
\end{figure}

The data obtained with XMM-Newton (EPIC MOS1, MOS2 and pn)
are summarized in Table~\ref{table:data}.  They were all retrieved
from the XMM-Newton archive. 
The data were reduced in a standard way, and 
temperature maps were computed following the procedure described by
Durret et al. (2005), with a main improvement which is the inclusion
of pn data to MOS1 and MOS2 data. The event files are rebinned with a
pixel size of $12.8 \times 12.8$ arcsec$^2$ and for each pixel in the
grid we compute the RMF and ARF matrixes and fit a \textsc{mekal}
plasma model using \textsc{xspec} version 11.2. We set a minimum count
number of 900 counts per pixel after background subtraction.
The hydrogen column density was always left free to vary (except for
Abell~1763).
The corresponding temperature maps are shown in Figs.~\ref{fig:fig1}
and \ref{fig:fig2}. Typical relative errors, $\sigma_{T}/T$, 
for each pixel are 10--15\% at $1 \sigma$ significance level.

\section{Discussion}

The temperature map of Abell~68 appears quite constant throughout the
cluster, with a mean temperature and metallicity of the X-ray gas of
$8.99 \pm 0.22$~keV and $0.23 \pm 0.03$ Z$_0$ for a hydrogen
absorption ${\rm N_H\ = 4.94\ 10^{20}\ cm^{-2}}$. A hotter region
observed towards the east edge of the map could possibly indicate a
merger coming in from this direction. Little is known about this
cluster, so we cannot reach firm conclusions on what is happening.

Abell~2390 has a mean temperature and metallicity of $9.46 \pm
0.20$~keV and $0.31 \pm 0.03$ Z$_0$ for ${\rm N_H\ = 6.80\ 10^{20}\
cm^{-2}}$.  The X-ray temperature of Abell~2390 is notably cooler in
the center, with a large and somewhat hotter zone in the north region
and a much hotter and more concentrated zone in the south, suggesting
the previous existence of one or two mergers in the north south
direction (possibly one coming from the north and one from the south).

Abell~1763 has a mean temperature and metallicity of $5.38 \pm
0.18$~keV and $0.24 \pm 0.04$ Z$_0$ for ${\rm N_H\ = 5.87\ 10^{20}\
cm^{-2}}$ (the absorption had to be fixed to its Galactic value,
otherwise the X-ray spectra were impossible to fit).  The X-ray
temperature map is quite uniform, with a hotter region 
($kT \approx 10.0 \pm 1.5$  keV) towards the
north west, suggesting a merger along the south east to north west
direction.

Abell~2744 is a complicated cluster undergoing several mergers, as
derived from XMM-Newton (Zhang et al. 2004, Finoguenov et al. 2005)
and optical (Boschin et al. 2006) data. Its X-ray gas has a mean
temperature and metallicity of $9.49 \pm 0.31$~keV and $0.21 \pm 0.04$
Z$_0$ for ${\rm N_H\ = 1.31\ 10^{20}\ cm^{-2}}$. The temperature map
reveals a hot elongated region crossing the cluster from north to
south, and of a second hot region towards the south east, and the gas
is rather hotter in the central region, in agreement with Finoguenov
et al. (2005). The two hotter regions can be interpreted in a
scenario of two successive major mergers, both coming from the south
east, the central one being observed after the core passage, as
suggested by Boschin et al. (2006). This agrees with the general
picture proposed by Kempner \& David (2004) but implies a more
advanced merging phase than that derived by these authors.

Finally, Cl~2137-2353, has a
mean temperature and metallicity of the X-ray gas of
$4.5 \pm 0.1$~keV and $0.37 \pm 0.04$ Z$_0$ (solar) for ${\rm N_H\ =
3.55\ 10^{20}\ cm^{-2}}$.  The temperature map reveals a very
homogeneous temperature throughout the cluster, except for a somewhat
hotter region towards the north west.  Therefore there is no clear
evidence for any type of merging in this object.

\section{Conclusions}

Even clusters with a relatively relaxed appearance can be undergoing,
or have undergone recently, one or several merging events, as revealed
by X-ray temperature maps.  Metallicity maps as well as optical
(imaging and spectroscopy) and radio data, together with the
comparison with numerical simulations, are now needed to disentangle
the effects of mergers and characterize better the history of
formation of these clusters.

\noindent
{\bf Acknowledgements}

We acknowledge financial support from the CAPES/COFECUB and CNES.

\end{document}